# Canvas Adoption: Assessment and Acceptance of the Learning Management System on a Web-Based Platform

**Julius G. Garcia**
Technological University of the Philippines
julius.tim.garcia@gmail.com

**Mark Gil T. Gañgan**
Isabela State University
mark28gil1991@gmail.com

**Marita N. Tolentino**
University of the East – Manila
marita.tolentino@ue.edu.ph

**Marc Ligas**
University of the East - Manila
marc.ligas@ue.edu.ph

**Shirley D. Moraga**
University of the East - Manila
shirley.moraga@ue.edu.ph

**Amelia A. Pasilan**
University of the East - Caloocan
amelia.pasilan@ue.edu.ph

## Abstract

*The acquisition of non-proprietary and proprietary learning management system has provided a richer learning experience to users and raised interest among education providers. This study aims to assess student adoption of Canvas as a new learning management system and its potential as a web-based platform in the e-learning programme of the University of the East. This study also assessed student readiness in using Canvas. A survey was administered to 214 students of the University of the East through snowball sampling. An Exploratory Factor Analysis was conducted to examine the validity of the model. A Confirmatory Factory Analysis was used to validate the Exploratory Factor Analysis results and analyse the correlation of the constructs. A Structural Equation Modelling was conducted to analyse the relationships between the constructs, which were evaluated using fit indices. Adopted from the Technology Acceptance Model, the constructs perceived ease of use, perceived usefulness, and attitude were studied. The study reveals that students' perceived usefulness and attitude towards using Canvas in a web-based platform have direct and significant effects on their intention to use Canvas. The students' perceived ease of use has a significant effect on their perceived usefulness but has no significant effects on their attitude towards the use of Canvas. The students' technological maturity and prior experience in using a learning management system*

*influenced their beliefs on the adaptation of similar technology. Exploring the potential benefits of Canvas and factors affecting the students' adoption amplifies access to quality education to fulfil educational directives. Furthermore, educational institutions should explore technological migration related to teaching and learning processes.*

# Introduction

Canvas is a collaborative platform for institutions, industry experts, teachers and students with more than 30 million users (IBL News, 2019) worldwide. The web- and cloud-based application provides newer features of the course management platform that include migration tools, a dashboard for learning outcomes, peer review and e-portfolios, real-time face-to-face communication tools such as video chat and screen sharing, and also an assessment tool that covers simple to complex grade point rubrics. Canvas also incorporates SpeedGrader, which is an application that accepts a variety of web-viewable formats for documents, blogs and videos and provides a specialised section for teachers to view and grade a student's coursework using the rubric as well as provide feedback.

In the adaptation of online educational programmes, colleges and universities have readily invested in such technological infrastructure. Canvas' new mobile features that are on a native cloud-based platform provide easy and comfortable access to learning materials while allowing students to collaborate with their peers, use it on the move and even take online examinations. This results in a more personalised learning journey for the students. Likewise, teachers have real-time access to data and can provide up-to-date assessment and grading results, which in turn provides a richer and more pleasant learning experience for both teachers and students.

Thus, the University of the East, Manila and Caloocan Campus is enhancing the adaptation of the latest innovations in teaching and learning ("Education Trailblazers," 2018). The continuous acquisition and upgrading of key enterprise systems and technology at the university provide sustainable value to the academic community. Specifically, the migration to a new learning management system (LMS) is one of the key technological considerations and solutions to fulfil institutional directives.

# Literature Review

With the rapid growth of web technology, web-based learning has become a significant area for research. With the provision of the Internet, portable devices like smartphones, laptops, and personal digital assistants allow instantaneous access to web-based applications and also learning environments. Thus, the desired information can be conveniently acquired on the web. Moreover, time and space constraints can be overcome to enhance self-directed learning in the process (Thuering et al., 1995).

Web-based learning platforms allow students to impart their ideas and discuss solutions. Navigating web resources allows students to construct knowledge and control knowledge construction and navigation processes (Kashihara & Hasegawa, 2005; Hasegawa & Kashihara, 2006). Also, teachers can assess their contributions and monitor their interactions through web-based learning facilities in a forthright approach (Andresen, 2009). Educational support resources on the web provide an avenue for teachers to encourage their students to explore.

The strategic vision of each institution is to enhance teaching and learning. Given the significant increase in the availability of online learning classes, universities are readily acquiring and adopting new LMSs. The acquisition of an LMS by higher education institutions has been a significant investment in information and communication technology in the past decades (Coates, James, & Baldwin, 2005; McGill & Klobas, 2009).

According to Basa ("Education Trailblazers," 2018), "Canvas allows us to develop courses that are geared to student-centred strategies. It pushed the envelope of teaching strategies development and helps us avoid being confined to the classroom in an enjoyable and engaging manner". With its user-friendliness, Canvas provides a simplified online learning experience for users, teachers and students even through the use of a mobile or tablet. The traditional classroom teacher can shift using the LMS to share and encourage students to discuss and give their input. According to Troy Martin ("Educational technology platform," 2016, para. 6), "teachers are beginning to move learning activities such as quizzes, assignment submissions and peer reviews from the classroom online, and are slowly extending classroom activities (such as discussions) outside the physical walls and into the space that the LMS provides."

The migration and adoption of a new LMS have driven the institution and its stakeholders to learn and investigate the new technology. Gaining insights from the adapted new technology is imperative to the pursuit of innovation and relevance in this digital age. The study of Such et al. (2017) described LMS migration as the stakeholders' movement to find a better state for institutional praxis and academic purposes at the university level. The results of these studies revealed that the involvement of stakeholders and faculty is a crucial factor. It also identified the four major factors that institutions should consider when migrating to LMS, i.e. (1) communication power, (2) catalyst for change, (3) technological compatibility, and (4) confidence or trust in the system, which needs the close involvement of faculty and stakeholders. The success of technology integration in teaching and learning depends on how teachers and students adapt to change and embrace the available technology (Pajo & Wallace, 2001). The foremost consideration in selecting and migrating to a new LMS should be greatly influenced by pedagogical considerations. The LMS selection must be anchored to exemplary teaching principles and good educational practices (Harrington et al., 2006; Chickering & Gamson, 1987). Otherwise, students may draw a negative perception of the learning management system, either perceiving it simply as a platform or tool to retrieve their documents or a medium to communicate with their teachers. This hardly justifies its cost and may potentially hamper system migration (Lonn & Teasley, 2009).

**Technology Acceptance Model**

The theory of reasoned action, which predicts individual behaviour based on their established attitudes and behavioural intentions, has paved the way to the development of the Technology Acceptance Model (TAM). Behavioural intention is the degree of strength or an individual's intention to accomplish a specific task, whether to adopt or utilise a system (Fishbein & Ajzen, 1975). Individual intention influences behaviour where actions are determined by normative, control and behavioural beliefs (Ajzen, 1991). Social norms and individual attitude towards a behaviour, i.e. behaviour approval or disapproval of an individual or group, are determinants or constructs of the intention.

In the context of TAM, attitude mediates the affective response between perceived usefulness and perceived ease of use to the behavioural intention of using the system. Attitude is an individual's desire to use a system (Karjaluoto et al., 2002). According to Fishbein and Ajzen (1975), there are two classifications of attitude, i.e. (1) attitude towards the object or system, and (2) attitude towards the behaviour. The overall attitude is a predecessor to the intention to adopt the system (Davis, 1989). It also proposes that intention is related to and directly influences the individual's behaviour in using the system (Davis et al., 1989).

Perceived usefulness is the degree to which an individual accepts that using a specific framework or modern innovation would improve his or her job performance (Davis, 1989; Davis et al., 1989; Davis, 1993). According to Davis et al. (1992), the individual's intrinsic motivation to perform a task is related to their perceived pleasure and satisfaction. Essentially, Mathwick et al. (2001) defined perceived usefulness as the degree to which a person deems a specific framework to boost his or her job performance. Fred Davis (1989) characterised perceived ease of use as the degree to which an individual accepts that employing a specific system would be free from effort. It is conceivable that instructive innovation through a high degree of perceived usefulness is expected to initiate positive states of mind. Moreover, it emphasises that perceived usefulness mediates the impact of perceived ease of use on the mental state (Moon & Kim, 2001). This gives a thought that perceived usefulness has a coordinate impact on demeanour, whereas perceived ease of use influences demeanour by implication through perceived usefulness.

A later survey found that these two factors received considerable consideration in an incredible number of earlier computer innovation acceptance and adoption and showed a significant impact in both coordinate (perceived usefulness) and circuitous (perceived ease of use) constructs on deliberate computer innovation utilisation (e.g. Legris et al., 2003). Moreover, recent studies and discourses showed the relationship of perceived usefulness and perceived ease of use as factors in predicting user's purpose to utilize computer and technological innovation (Ma, Gam, & Banning, 2017; Falode, 2018; Denaputri & Usman, 2019; Wicaksono & Maharani, 2020). In this regard, this study explores the direct effects of the attitude to behavioural intention and the mediating effects of perceived usefulness and perceived ease of use to attitude towards the use of the system.

**Figure 1**

*Research Framework of the Study*

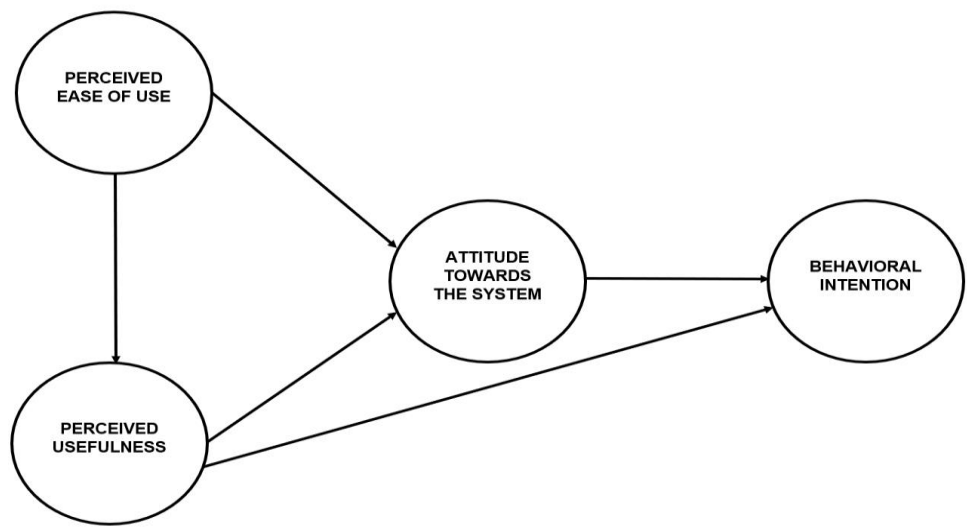

**Problem Statement**

Therefore, this concept leads to the proposed research framework as illustrated in Figure 1, which provides the measures for the adoption of the Canvas LMS. Specifically, this study sought to answer the following questions:

   i. What is the student readiness in using Canvas as a web-based learning platform?
   ii. What is the user's perceived acceptability of Canvas as a web-based learning platform?
   iii. Do perceived ease of use and perceived usefulness have a significant effect on the attitude towards using Canvas as a web-based LMS?
   iv. Do attitude and perceived usefulness have a significant effect on the behavioural intention to use Canvas as a web-based LMS?

## Research Objectives

In this regard, this study aims to assess student adoption of Canvas as a new LMS and its potential as a web-based platform in the e-learning programme of the University of the East. It also aims to assess student readiness in using Canvas as a web-based learning platform. Hence, this study has the following specific objectives:

   i. To examine the students' extent of use of personal computing devices and the Internet.
   ii. To determine the students' level of proficiency in mobile and web applications.
   iii. To investigate the factors that influence the students' adoption of the new LMS.
   iv. To recommend measures to improve the adoption of the new LMS to stakeholders.

## Research Hypothesis

$H_i1$: Attitude towards the use of the system has a direct significant effect on the behavioural intention to use Canvas as a web-based LMS.
$H_i2$: Perceive usefulness has a direct significant effect on the behavioural intention to use Canvas as a web-based LMS.
$H_i3$: Perceive usefulness has a direct significant effect on the attitude towards using Canvas as a web-based LMS.
$H_i4$: Perceived ease of use has a direct significant effect on the attitude towards using Canvas as a web-based LMS.
$H_i5$: Perceived ease of use has a direct significant effect on the perceived usefulness of Canvas as a web-based LMS.

## Research Method

**Participants and Localities**

There were 214 valid responses (Male = 103; Female = 111; Mean age = 19.53) collected from senior high school students (n=103) and college students (n=111) of the University of the East. The questionnaire is used to assess the students' behavioural intention to use Canvas as a web-based learning management system. Most of the respondents are Filipinos (n=210) and the rest are foreigners (n=4), each of whom is Japanese, Iranian, English and Pakistani, respectively.

**Data Collection**

An online and paper-based survey was provided to 214 students of the University of the East during the pre-implementation of the new LMS from Moodle to Canvas. Only two departments were part of the pre-implementation process for the new LMS, i.e. the College of Computer Studies and the Senior High School Department. The respondents were randomly selected through snowball sampling. The teachers from selected departments administered the survey to the selected college students endorsed by their colleagues.

**Data Measurement**

The questionnaire was adopted from TAM. The questionnaire was composed of 1) Demographics; 2) Personal Computing Devices; 3) Extent of Device Use; 4) Level of Proficiency; 5) Perceived Ease of Use; 6) Perceived Usefulness 7) Attitude, and 8) Behavioural Intention to Use. We adopted a validated scale to develop our survey questionnaire, employing a five-point scale (1 = 'strongly disagree' to 5 = 'strongly agree' for the level of agreement, and 1 = 'never' to 5 = 'always' for the extent or frequency). To test the reliability of each item, Cronbach's alpha analysis was performed to ensure the alpha measurement α >.60 acceptability for TAM (Krishnaveni & Meenakumari, 2010). The focus of this study is the attitude towards behaviour in which the following mediating constructs, i.e. perceived ease of use and perceived usefulness, influenced attitude towards using Canvas.

**Data Analysis**

Descriptive statistical analysis was conducted to determine the mean and standard deviation of the constructs and items using SPSS. An exploratory factor analysis (EFA) was performed to identify the correlation between the variables. A separate dataset was used for exploratory factor analysis. Subsequently, the rest of the data were included in the dataset and utilised for confirmatory factor analysis (CFA). Then, the CFA was conducted to validate the EFA results and also to test the data fitness to the research model using IBM AMOS. Structural Equation Modelling (SEM) was performed to test the designed model subjected to fit indices. Moreover, a path analysis was conducted to identify the relationships between the constructs.

# Findings

As illustrated in Figure 2, out of 214 respondents, 97.66% have smartphones, 76.64 own laptop computers, 50% have desktop computers at home, 33.18% own tablets and only 5.61% of the respondents have Mac devices. The low acquisition and ownership of Mac devices can be attributed to its price. Moreover, most schools in the Philippines use Windows-based computers.

**Figure 2**

*Personal Computing Devices*

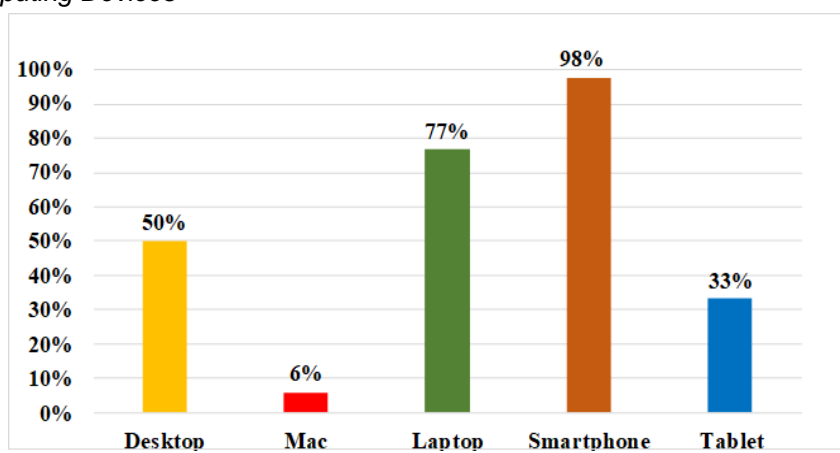

Among the devices shown in Table 1, smartphones (M=4.60, 0.75) are always utilised by the students because of their physical mobility and communication features. Laptops (M=3.30, 1.27) and desktops (M=2.78, 1.40) are occasionally utilised. Both Mac devices (M=1.88, 0.86) and tablets (M=2.04, 1.28) are almost never utilised by the students.

**Table 1**

*The extent of Use of Personal Computing Devices*

| Personal Computing Devices | Mean | SD | Verbal Interpretation |
|---|---|---|---|
| Desktop | 2.78 | 1.40 | Occasionally/Sometimes |
| Mac | 1.88 | 0.86 | Almost never |
| Laptop | 3.30 | 1.27 | Occasionally/Sometimes |
| Smartphone | 4.60 | 0.75 | Every time |
| Tablet | 2.04 | 1.28 | Almost never |

*Note.* 1.00-1.79 = Never, 1.80-2.59 = Almost never, 2.60-3.39 = Occasionally/Sometimes, 3.40-4.19 = Almost every time, 4.20-5.00 = Every time

As shown on Table 2, students use the Internet at home (M = 4.64, SD = 0.63). They occasionally or sometimes use the Internet at school (M = 2.84, SD = 1.53). However, while outside (M = 3.66, SD = 1.11), either school or home, they use the Internet almost every time.

**Table 2**

*Respondents' Extent of Use of the Internet*

| The extent of Use of the Internet at: | Mean | SD | Verbal Interpretation |
|---|---|---|---|
| Home | 4.64 | 0.63 | Every time |
| School | 2.84 | 1.53 | Occasionally/Sometimes |
| Outside (Travelling, etc.) | 3.66 | 1.11 | Almost every time |

*Note.* 1.00-1.79 = Never, 1.80-2.59 = Almost never, 2.60-3.39 = Occasionally/Sometimes, 3.40-4.19 = Almost every time, 4.20-5.00 = Every time

In terms of using the Internet, the students' level of proficiency is very good (M = 3.93, SD = 0.86). In addition, they have very good proficiency in using Mobile/PC applications (M = 4.09, SD = 0.86) and web-based applications (M = 3.90, SD = 0.88), as indicated on Table 3.

**Table 3**

*Respondents' Level of Proficiency*

| Level of Proficiency | Mean | SD | Verbal Interpretation |
|---|---|---|---|
| Internet Use | 3.93 | .86 | Very Good |
| Mobile/PC Applications | 4.09 | .87 | Very Good |
| Web-Based Applications | 3.90 | .88 | Very Good |

*Note.* 1.00-1.79 = Poor, 1.80-2.59 = Fair, 2.60-3.39 = Good, 3.40-4.19 = Very good, 4.20-5.00 = Excellent

Among the constructs shown in Table 4, the perceived usefulness of Canvas obtained the highest mean (M = 3.93, SD = 0.75). Students believe that Canvas is useful in their learning process. Students also believe that Canvas can help them to quickly accomplish their coursework tasks. Although the difference between the latter constructs is 0.05, the attitude towards using Canvas received the second-highest mean (M = 3.88, SD = 0.73). This suggests that the students had a pleasant experience working on their coursework using the web-based Canvas. In general, the students agree that Canvas can help them improve their learning.

**Table 4**

*Students' Acceptability Rating on Canvas as a Web-based LMS*

| Constructs | Mean | SD |
|---|---|---|
| Perceived Ease of Use of Canvas (PEOU) | 3.67 | 0.64 |
| Perceived Usefulness of Canvas (PU) | 3.93 | 0.75 |
| Attitude towards Using Canvas (AT) | 3.88 | 0.73 |
| Behavioural Intention to Use of Canvas (BI) | 3.72 | 0.84 |

*Note.* 1.00-1.79 = Strongly disagree, 1.80-2.59 = Disagree, 2.60-3.39 = Neutral, 3.40-4.19 = Agree, 4.20-5.00 = Strongly agree

**Model Validation and Testing**

As shown in Table 5, the result of the Maximum Likelihood extraction using Varimax rotation reveals that the factor loadings of each item are valid. The items of each construct, i.e. PEOU, PU, AT, and BI, appeared in a single dimension and yielded a range of factor loading results of .605 to .814 with a relatively high variance of 78.75%, which determines dimensionality of measurements among the items of the constructs. Based on the EFA results, the four constructs fall on the four factors or dimensions that are deemed appropriate to the current study.

**Table 5**

*Factor Loadings and Correlation Matrix*

| Items | FACTORS | | | |
|---|---|---|---|---|
| | 1 | 2 | 3 | 4 |
| PU3 | **.814** | .306 | .205 | .150 |
| PU2 | **.804** | .253 | .190 | .161 |
| PU1 | **.762** | .263 | .325 | .140 |
| BI1 | .303 | **.797** | .210 | .164 |
| BI2 | .225 | **.786** | .171 | .098 |
| BI3 | .218 | **.780** | .230 | .078 |
| AT1 | .238 | .195 | **.793** | .165 |
| AT2 | .236 | .242 | **.746** | .047 |
| AT3 | .118 | .129 | **.656** | .127 |
| PEOU2 | -.013 | .134 | .003 | **.790** |
| PEOU3 | .234 | .077 | .144 | **.767** |
| PE3 | .131 | .050 | .159 | **.605** |

Each construct in Table 6 demonstrates an acceptable level of reliability coefficient or internal consistency while taking into account that a reliability coefficient of .70 or higher is generally considered acceptable. The Cronbach's Alpha values are α = .78 for PEOU, α =.92 for PU, α =.83 for AT, and α = .89 for BI, respectively, which confirms their reliability. In addition, the composite reliability of the constructs ranged from 0.78 to .92, which are higher than the acceptable recommend value of .70 (Hair et al., 1998). This indicates that all constructs in the measurement model provided evidence of good reliable measures. In terms of convergent validity, the data shows the average variance extracted values ranged from .55 to .78, which suggest that the constructs' overall convergent validity were above the .5 threshold needed to suggest a reasonable overall convergent validity (Fornell and Larcker, 1981). In addition, the constructs' composite reliability ranges from 0.78 to 0.92, which are higher than the acceptable recommend value of .70 (Hair et al., 1998). This strongly indicates that the constructs in the model provided evidence of reliability.

**Table 6**

*Internal Factor Reliability: Cronbach's Alpha, Composite Reliability (CR), Average Variance Extracted (AVE) and Maximum Shared Variance (MSV)*

| Constructs | Cronbach's Alpha | CR | AVE | MSV | MaxR(H) | PU | BI | AT | PEOU |
|---|---|---|---|---|---|---|---|---|---|
| PU | 0.92 | 0.915 | 0.783 | 0.411 | 0.916 | 0.885 | | | |
| BI | 0.89 | 0.893 | 0.735 | 0.411 | 0.898 | 0.641*** | 0.857 | | |
| AT | 0.83 | 0.832 | 0.625 | 0.345 | 0.855 | 0.587*** | 0.542*** | 0.79 | |
| PEOU | 0.78 | 0.781 | 0.547 | 0.171 | 0.824 | 0.413*** | 0.327*** | 0.350*** | 0.74 |

*Note.* * $p < 0.050$, ** $p < 0.010$, *** $p < 0.001$

**Table 7**

*Confirmatory Factor Analysis*

| Constructs | Items | Latent standardized parameter | Measurement errors θ | t-value |
|---|---|---|---|---|
| Perceived Ease of Use | 1. Canvas would be easy to use using my desktop/laptop browser. | 0.63 | 0.035 | 7.656*** |
| | 2. It would be easy to access course materials in Canvas using my desktop/laptop browser. | 0.69 | 0.038 | 3.267*** |
| | 3. Canvas is easy to work around and operate using my desktop/laptop browser. | 0.87 | 0.043 | 8.597*** |
| Perceived Usefulness | 1. Using my desktop/laptop browser, Canvas would improve my ability to learn. | 0.88 | 0.018 | 6.232*** |
| | 2. Using my desktop/laptop browser, Canvas would help me accomplish my task quickly. | 0.87 | 0.019 | 7.416*** |
| | 3. Using my desktop/laptop browser, Canvas is useful for learning. | 0.90 | 0.019 | 7.053*** |
| Attitude | 1. I would like my coursework more if I used Canvas through my desktop/laptop browser. | 0.87 | 0.029 | 4.975*** |
| | 2. Canvas would be a pleasant experience if I use my desktop/laptop browser. | 0.82 | 0.028 | 6.582*** |
| | 3. It is a good idea to use Canvas in a desktop/laptop browser. | 0.67 | 0.035 | 8.92*** |
| Behavioural Intention | 1. I intend to continue using Canvas for classroom instruction and activities. | 0.90 | 0.026 | 5.566*** |
| | 2. It is likely that I will use Canvas in the future. | 0.83 | 0.03 | 7.628*** |
| | 3. When I need to do coursework activities, I prefer using Canvas for classroom instruction and activities in the future. | 0.84 | 0.032 | 7.406*** |

*Note.* * $p < 0.050$, ** $p < 0.010$, *** $p < 0.001$

After the CFA, 12 items were generated with three items under each dimension, as shown in Table 7 and illustrated in Figure 3. The model was evaluated using the following model fit index criteria; Chi square statistics, Normed fit index (NFI), Goodness of fit index (GFI), Root mean square Residual (RMSR), Comparative fit index (CFI), Incremental fit index (IFI) and Root Mean Square Error of Approximation (RMSEA). Results revealed that the model has overall goodness of fit to the data. It also demonstrated a good explanatory power for this study, as indicated in Table 8.

**Table 8**

*Summary of the Model Fit Indices' Results*

| Fit Indices | Index Value | Acceptable Value |
|---|---|---|
| $X^2$/DF | 1.482 | <3 |
| Root Mean Square Residual (RMSR) | 0.024 | ≤ 0.1 |
| Goodness of fit index (GFI) | 0.948 | ≥ 0.9 |
| Normed fit index (NFI) | 0.953 | ≥ 0.9 |
| Incremental fit index (IFI) | 0.984 | ≥ 0.9 |
| Comparative fit index (CFI) | 0.984 | ≥ 0.9 |
| Root Mean Square Error of Approximation (RMSEA) | 0.048 | < 0.6 |

**Figure 3**

*Canvas Adoption on Web-based Platform Interrelationship Model*

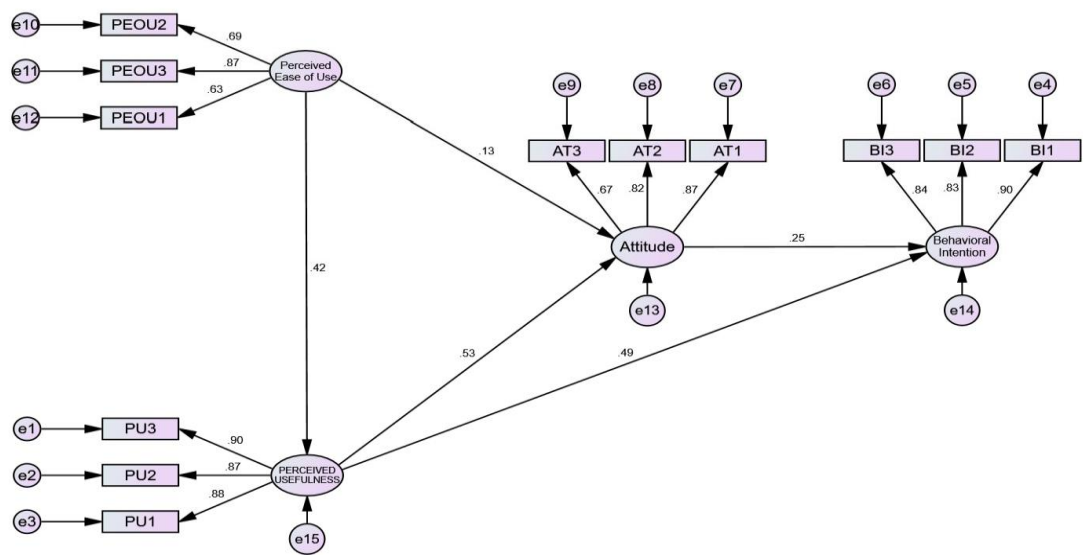

The summary of the path coefficients illustrated in Figure 4 shows the direct and indirect effects or contribution among the constructs. The perceived ease of use of Canvas has an effect on perceived usefulness of Canvas (*β = 0.42, p<0.001*) but has no effect on attitude towards Canvas. This explains that 17% of its total variance accounted for its indirect effect on the attitude towards Canvas and behavioural intention when moderated by perceived usefulness.

Perceived usefulness has a direct and significant effect on the behavioural intention to use Canvas (*β = 0.49, p<0.001*). Perceived usefulness also has a significant effect on the attitude towards Canvas (*β = 0.53, p<0.001*), which is mediated by behavioural intention. This explains that 36% of the total variance accounted for behavioural intention. Lastly, attitude towards the system has an effect on the behavioural intention to use Canvas ($\beta = 0.25, p<0.05$).

**Model Elaboration**

**Figure 4**

*Structural Model Analysis Results*

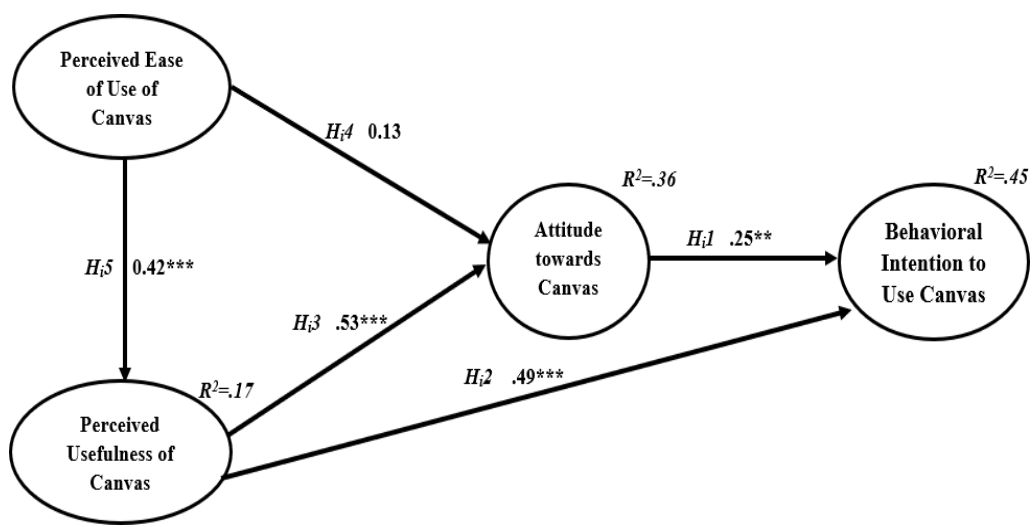

**Table 9**

*Summary Results of the Path Coefficients and Hypothesis*

| Path of Constructs | | | β | t-value | $R^2$ | Results |
|---|---|---|---|---|---|---|
| Behavioural Intention | <--- | Attitude towards the system | 0.25 | 3.10*** | 0.45 | $H_i1$ - Supported |
| Behavioural Intention | <--- | Perceived Usefulness | 0.49 | 6.06*** | | $H_i2$ - Supported |
| Attitude towards the System | <--- | Perceived Usefulness | 0.53 | 6.76*** | 0.36 | $H_i3$ - Supported |
| Attitude towards the System | <--- | Perceived Ease of Use | 0.13 | 1.67 | | $H_i4$ - Not Supported |
| Perceived Usefulness | <--- | Perceived Ease of Use | 0.42 | 5.19*** | 0.17 | $H_i5$ - Supported |

*Note.* * p < 0.050, ** p < 0.010, *** p < 0.001

Overall, the model explains 45% of the total variance in the behavioural intention to use Canvas. Based on the path analysis shown in Table 9, the alternative hypotheses $H_i1$, $H_i2$, $H_i3$ and $H_i5$, are supported, suggesting that attitude towards the system and its perceived usefulness yield a significant effect on the students' behavioural intention to use Canvas. Similarly, the students' perceived ease of use yields a significant effect on attitude and perceived usefulness. However, for $H_i4$, there is no support for the alternative hypothesis. This suggests the students' perceived ease of use of Canvas has no significant influence on their attitude to use Canvas.

# Discussion

The use of technological tools such as laptops, smartphones, tablets and desktop computers has influenced the teaching process and most importantly, student learning. Based on this study, smartphones and laptops are greatly utilised by the students. The utilisation of mobile devices can be a medium for continuous learning and a mobility tool for students. Likewise, it is more beneficial for students to access the LMS, especially for those who have both laptops and desktops, and moderate learning in the absence of either device. The students' extent of Internet use was determined to identify its potential contribution to the access and engagement of the students in the Canvas LMS. This study reveals that students are highly utilising the Internet at home. Therefore, it will provide teachers with the opportunity to expand learning and engage students with their coursework not only in the school but also at home through proper monitoring, intervention and motivation. Generally, the students have a very good level of proficiency in using mobile and web-based application. Combined with their very good level of proficiency in using the Internet, this may increase their level of understanding and operational skills in using Canvas on a web-based platform. Sometimes, coursework components can be overwhelming. The students' familiarity with the previous web-based LMS may contribute to the ease in operating and working around with the new LMS.

Moreover, the results show that the students' perceived usefulness of Canvas as a web-based LMS significantly contributes to their attitude towards the use of the system. This reveals that the students attributed their learning improvement and accomplishment with coursework and tasks to Canvas, which in turn influences their desire to utilise it. Likewise, the students' perceived ease of use of Canvas has a significant impact on the perceived usefulness of Canvas. This could explain that the features, tools, functionalities, interface design and the learning environment within Canvas contributed to the students' learning improvement and development. However, the students' perceived ease of use has no significant effect on the attitude towards the use of Canvas. This result corresponds to previous studies that indicated that perceived ease of use is less likely to or has little effect on those with no experience or has no effect on those with experience on their attitude and behavioural intention (Pikkarainen et al., 2004; Hu et al., 1999; Wu & Wang, 2005; Venkatesh & Davis, 1996; Venkatesh, 2000; Koufaris, 2002). This can be attributed to the students' prior experience in using an LMS that has transformed and change their perception of ease of technology use over time. The students' technological maturity in using the LMS has increased, which could reduce its direct effect on their attitude and intention towards the use of Canvas.

## Conclusion

This study assesses the students' adoption of Canvas and its potential as a web-based platform in the e-learning programme of the University of the East. Students perceived Canvas as an important medium to improve their learning because it can assist them with their tasks, has functionalities that are easy to operate, and also provides ease of access to their coursework. Canvas provides a pleasant experience to students, which leads to their intention to use and adopt the new LMS. In light of the results, it is perceivable that students will utilise Canvas as a web-based platform in their learning as well as academic tasks. Because this study was conducted during a migratory or pre-implementation phase, it is crucial to provide technical assistance and support to students in utilising Canvas within and beyond the classroom. This could allow students to create and innovate new ideas through the new LMS or similar technology. This could also equip them with essential skills and confidence in using technology. Teachers should also promote the use of online learning platforms and explain to students their technological benefits in learning. Considering that the web-based learning management platform has been useful in the teaching-learning process, teachers should also promote the use of corresponding mobile-based learning applications that are downloadable at Google Play Store and Apple App Store. Moreover, it is imperative to investigate the effects of adoption and migration among teachers and administrators. Since this study was conducted during the pre-implementation phase with limited sample size, it is recommended that a future survey increases the population size and explores other variables such as social influence, satisfaction and other underlying social and cultural factors to improve the variance in this study. Originally, moderating factors such as profile, education, age and gender were included in this study but were removed because they yielded insignificant effects to the constructs. However, these factors should also be considered with relatively large sample size. Moreover, the study was conducted in a private university with students at the secondary and tertiary level. Therefore, it is also recommended that a similar study is conducted in public schools at different levels of study.